\documentclass[%
 aps, %aps
 amsmath,amssymb,
preprint,%
]{revtex4-2}

\usepackage{graphicx}% Include figure files
\usepackage{dcolumn}% Align table columns on decimal point
\usepackage{bm}% bold math
\usepackage[utf8]{inputenc}
\usepackage[T1]{fontenc}
\usepackage{mathptmx}
\usepackage{etoolbox}
\usepackage{color}
\usepackage{lineno}

\makeatletter
\def\@email#1#2{%
 \endgroup
 \patchcmd{\titleblock@produce}
  {\frontmatter@RRAPformat}
  {\frontmatter@RRAPformat{\produce@RRAP{*#1\href{mailto:#2}{#2}}}\frontmatter@RRAPformat}
  {}{}
}%
\makeatother

\makeatletter
\DeclareRobustCommand{\iscircle}{\mathord{\mathpalette\is@circle\relax}}
\newcommand\is@circle[2]{%
  \begingroup
  \sbox\z@{\raisebox{\depth}{$\m@th#1\bigcirc$}}%
  \sbox\tw@{$#1\square$}%
  \resizebox{!}{\ht\tw@}{\usebox{\z@}}%
  \endgroup
}
\makeatother
\begin{document}

%\preprint{AIP/123-QED}
%\linenumbers
\title[Influence of the magneto-volume effect on the transient reflectivity of MnSi]{Influence of the magneto-volume effect on the transient reflectivity of MnSi}
% Impact of the ultrafast magneto-volume effect on the transient reflectivity of the chiral magnet MnSi
% Exploring the 
\author{J. Kalin$^*$}\email{jantje.kalin@ptb.de}
\affiliation{Physikalisch-Technische Bundesanstalt, 38116 Braunschweig, Germany}

\author{S. Sievers}%
\affiliation{Physikalisch-Technische Bundesanstalt, 38116 Braunschweig, Germany}

\author{A. Bauer}
\affiliation{Physik-Department, Technische Universität München, 85748 Garching, Germany}
\affiliation{Zentrum für QuantumEngineering (ZQE), Technische Universität München, 85748 Garching, Germany}

\author{H. W. Schumacher}
\affiliation{Physikalisch-Technische Bundesanstalt, 38116 Braunschweig, Germany}

\author{H. Füser}
\affiliation{Physikalisch-Technische Bundesanstalt, 38116 Braunschweig, Germany}

\author{C. Pfleiderer}
\affiliation{Physik-Department, Technische Universität München, 85748 Garching, Germany}
\affiliation{Zentrum für QuantumEngineering (ZQE), Technische Universität München, 85748 Garching, Germany}
\affiliation{Munich Center for Quantum Science and Technology (MCQST), Technische Universität München, 85748 Garching, Germany}
\affiliation{Heinz Maier-Leibnitz Zentrum (MLZ), Technische Universität München, 85748 Garching, Germany}

\author{M. Bieler}
\affiliation{Physikalisch-Technische Bundesanstalt, 38116 Braunschweig, Germany}

\date{\today}% It is always \today, today,

\begin{abstract}
%Transient reflectivity measurements are an invaluable tool to study ultrafast thermal processes and contribute to the understanding of electron-lattice interactions and thermal diffusion, providing key information for the numerical modeling of thermally-excited ultrafast magnetic phenomena. Accurately extracting the material's temperature response from transient reflectivity requires a precise understanding of ultrafast processes influencing reflectivity, which is challenging in magnetic materials due to the intricate interplay of electron, charge, and spin degree of freedom.
The magneto-volume effect is a well-established yet frequently overlooked phenomenon in magnetic materials that may affect a wide range of physical properties. Our study explores the influence of the magneto-volume effect on the transient reflectivity of MnSi, a well-known chiral magnet with strong magnetoelastic coupling. We observe a unipolar reflectivity transient in the paramagnetic phase, contrasting with a bipolar response in phases with magnetic long-range order. Comparing our findings with thermal expansion from literature, we establish that the bipolar response originates in the magneto-volume effect which dominates the thermal expansion and influences the optical reflectivity. Our results highlight not only that the magneto-volume effect must be considered when discussing transient reflectivity measurements of magnetic materials but also that such measurements permit to study the characteristic time scales of the magneto-volume effect itself, contributing to a deeper understanding of this often-neglected phenomenon.

%The magneto-volume effect is a well-established yet frequently overlooked phenomenon in magnetic materials due to its inherent theoretical complexity.
%This study investigates the influence of magneto-volume effect on the transient reflectivity of magnetic materials.} Herein, we report transient reflectivity measurements on MnSi, an archetypical chiral magnet with strong magnetoelastic coupling. We find a unipolar reflectivity transient in the paramagnetic state and a bipolar response in states exhibiting magnetic long-range order. Comparison with thermal expansion data from literature shows that the bipolar response originates in the magneto-volume effect dominating the thermal expansion and influencing the optical reflectivity. \textcolor{blue}{These findings highlight that magneto-volume effect needs to be taken into account for accurately interpreting the transient reflectivity of magnetic materials. Furthermore, they open the possibility to investigate the dynamical behavior and time constants of the magneto-volume effect by transient reflectivity measurements, which contributes to a better understanding of this often neglected phenomenon}.
\end{abstract}

\maketitle
The interaction of femtosecond laser pulses with magnetic materials gives rise to intriguing phenomena, such as ultrafast demagnetization \cite{koopmans2010explaining, beaurepaire1996ultrafast}, all-optical switching \cite{stanciu2007all} and excitation of collective spin dynamics \cite{van2002all}. To effectively model these effects in micromagnetic simulations and investigate the underlying physical processes, knowledge of the transient temperature response of the material after thermal excitation represents a key prerequisite \cite{kalashnikova2016exchange, atxitia2007micromagnetic}.
%Understanding the transient temperature response of materials subjected to femtosecond laser pulses is crucial for effectively modeling phenomena like ultrafast demagnetization \cite{koopmans2010explaining, beaurepaire1996ultrafast}, all-optical switching  \cite{}, and laser-excited magnetization dynamics \cite{van2002all} in micromagnetic simulations and by that investigate the underlying physical processes.
The standard technique for determining the transient temperature responses is based on time-resolved transient reflectivity measurements, probing the change of the optical reflectivity $\Delta R$ due to thermal laser excitation.
 
In the typical framework of transient reflectivity \cite{allen1987theory,winsemius1976temperature, liu2021differentiating, hohlfeld2000electron,schoenlein1987femtosecond, djordjevic2006comprehensive}, the reflectivity of a metal is used as a thermometer, as $\Delta R$ is assumed to be proportional to the transient electron \cite{allen1987theory,schoenlein1987femtosecond, eesley1983observation} and lattice \cite{winsemius1976temperature, djordjevic2006comprehensive} temperature. The evolution of these temperatures as a function of time is described by the two-temperature model \cite{anisimov1974electron}, which treats the electronic and lattice systems as coupled heat baths. Thus, the laser-induced changes of the optical reflectivity are commonly described as the result of two contributions \cite{liu2021differentiating}: (i) thermal alteration of the electron density distribution \cite{allen1987theory, hohlfeld2000electron} and (ii) excitation of phonons associated with thermal expansion \cite{eesley1983observation, winsemius1976temperature, djordjevic2006comprehensive}.

However, in magnetic materials additional effects may impact the transient reflectivity due to the quenching of magnetic order upon femtosecond laser excitation, challenging the standard picture of transient reflectivity. Perhaps most notably, the magneto-volume effect, i.e., the volume change arising from a change of the magnetic order \cite{joule1847xvii} due to magnetoelastic coupling, gives rise to modifications of the electron interband and intraband transition rates at optical frequencies and thus to changes of the optical reflectivity. The magneto-volume effect was extensively studied in thermal equilibrium \cite{joule1847xvii,mckeehan1926magnetostriction, matsunaga1982magneto, fawcett1970magnetoelastic, stishov2007magnetic, petrova2016thermal}, but its impact on ultrafast timescales \cite{schmising2008ultrafast, reid2018beyond, kubota1975transient} and its consequences on optical reflectivity is less explored. %Yet, understanding the influence of quenched magnetic order and the magneto-volume effect on the optical reflectivity is crucial for accurately determining the transient temperature response of the material after laser excitation in transient reflectivity measurements.

%% In this letter
In this article, we investigate the influence of long-range magnetic order and the magneto-volume effect on the transient reflectivity of the archetypical chiral magnet MnSi in optical pump-probe experiments. We confirm the alteration of magnetic order upon laser excitation in time-resolved magneto-optical Kerr effect (TR-MOKE) measurements and compare the magnetization dynamics to the transient reflectivity. Our study reveals, that the temporal evolution of the transient reflectivity changes drastically from a unipolar behavior in the paramagnetic phase to a bipolar signal in the states with long-range magnetic order.
By a comparison of the transient reflectivity and thermal expansion we show that the bipolar signal is caused by the magneto-volume effect dominating the thermal expansion. These results contribute to an improved understanding of transient reflectivity of materials exhibiting magneto-volume effect, which is the case for most magnetic materials close to phase transitions. Further, this study shows how to accurately determine the transient temperature response of these materials.

\begin{figure}
\includegraphics{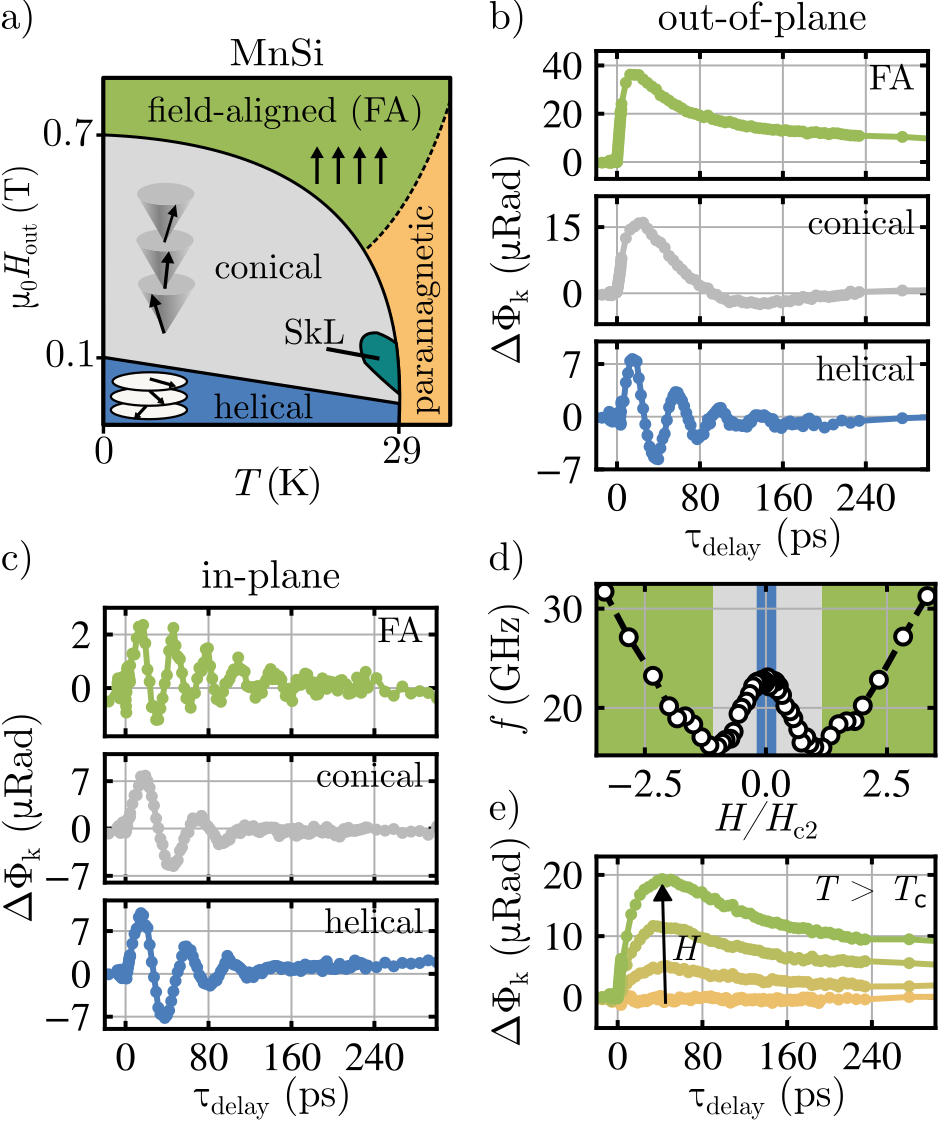}
\caption{\label{fig:Figure1} (a) Schematic magnetic phase diagram of MnSi for magnetic fields along <110>. The critical fields $\mu_0{H_\text{c1}\approx0.1\,\text{T}}$ and $\mu_0{H_\text{c2}\approx0.7\,\text{T}}$ for magnetic field transitions as well as the critical temperature ${T_\text{c}=29\,\text{K}}$ are indicated. (b) and (c) Typical out-of-plane (b) and in-plane (c) TR-MOKE signals in the field-aligned (FA), conical and helical phases at $10\,\text{K}$. (d) Dispersion extracted from the in-plane TR-MOKE signals at $10\,\text{K}$ indicating the characteristic spin modes of the helical, conical, and field-aligned phase. (e) Out-of-plane TR-MOKE signal of the paramagnetic phase for various magnetic fields ${[0, 0.2, 0.5, 1.0]\, \text{T}}$ at $39\,\text{K}$.     
 }
\end{figure}

%% sample % 
%MnSi is the ideal candidate for studying the influence of quenched long-range magnetic order and ultr
We focus on the bulk chiral magnet MnSi, owing to its pronounced magneto-volume effect \cite{matsunaga1982magneto,petrova2016thermal,stishov2008heat,fawcett1970magnetoelastic,stishov2007magnetic} and its magnetic phase diagram \cite{bauer2012magnetic}, depicted in Fig. \ref{fig:Figure1}(a), which permits to study the transient reflectivity of different magnetic states. At high temperatures, MnSi is a paramagnet. As the temperature is lowered beneath the critical temperature ${T_\text{c} = 29\,\text{K}}$, long-range magnetic order emerges. In this temperature regime, MnSi shows a helical phase for magnetic fields smaller than a critical field value $H_\text{c1}$. This phase is characterized by long-wavelength spin spirals, which are aligned along the magnetically easy axes <111> of the lattice. For magnetic fields above $H_\text{c1}$, the spins tilt towards the magnetic field direction, resulting in a conical phase. Finally, for magnetic fields larger than the critical field $H_\text{c2}$, the spins align collinearly in the field-aligned phase. Just below the critical temperature in a small regime in the magnetic phase diagram, MnSi shows a skyrmion lattice state with a spin swirling structure \cite{muehlbauer2009skyrmion}.

MnSi exhibits a significant magneto-volume effect \cite{matsunaga1982magneto,petrova2016thermal,stishov2008heat,fawcett1970magnetoelastic,stishov2007magnetic}, driven by magnetoelastic coupling. Due to that the sample's volume changes in response to variations in magnetic field and temperature, as the magnetic order variates with these parameters. Via the temperature dependence, the magneto-volume effect influences the thermal expansion and leads to an anomalous behavior.
For $T>>T_c$ and at low magnetic fields in the paramagnetic phase, MnSi shows an ordinary positive thermal expansion caused by the excitation of phonons and electrons, leading to a contraction of the lattice upon cooling. In contrast, the lattice of MnSi expands upon cooling for temperatures below and in proximity to $T_\text{c}$ and shows a negative thermal expansion. This behavior is explained by the magneto-volume effect surpassing the positive thermal expansion caused by phonons and electrons. Thereby, the magneto-volume effect is largest close to the phase transition at $T_\text{c}$.
At high magnetic fields above $T_\text{c}$, where the magnetic moments orient along the direction of the field, the magneto-volume effect and a negative thermal expansion is again observed \cite{stishov2008heat}. 

In this study, TR-MOKE and transient reflectivity measurements were performed simultaneously in a femtosecond pump-probe experiment. Laser pulses of about $150\,\text{fs}$ duration with a center wavelength of $800\,\text{nm}$ and a repetition rate of $76\,\text{MHz}$ are generated by a Ti:Sapphire oscillator. The linearly-polarized laser light is split into a pump and probe beam. The pump beam, which hits the sample under nearly normal incidence with a fluence of $2\,\mu\text{Jcm}^{-2}$, thermally excites the MnSi sample ($2\times2\times0.5\,\text{mm}^3$) by laser heating. In addition to steady-state heating, which is around $4\,\text{K}$ and accounted for when estimating the base temperature, laser heating with the femtosecond laser pulses introduces a temperature increase on the ultrafast timescale. %After the thermalization of non-thermal electrons and phonons, the three temperature model \cite{supplement} can be applied to estimate the maximum change in electron and spin temperature. 
At $10\,\text{K}$ and for the used laser fluence, laser heating in our experiment \cite{supplement} leads to a maximum change in electron, lattice and spin temperature of $\Delta T_\text{e,max}=18\,\text{K}$, $\Delta T_\text{L,max}=14\,\text{K}$ and $\Delta T_\text{s,max}=(15\pm3)\,\text{K}$, respectively.
The transient temperature increase modulates the optical properties of the sample on an ultrafast timescale and excites spin dynamics. By analyzing the probe beam after nearly normal reflection from the sample surface, the transient reflectivity is inferred from the thermally induced amplitude changes of the reflectivity $\Delta R/R$. In contrast, in TR-MOKE the change of the polarization state of the reflected probe beam is measured and quantified by the variations in the Kerr angle $\Delta \phi_\text{k}$. Due to the small incidence angle of the probe beam, $\Delta \phi_\text{k}$ is mainly sensitive to the changes of the out-of-plane magnetization component.  In order to enhance the detection sensitivity, the amplitude of the pump beam is modulated at a frequency of $3\,\text{kHz}$ by an optical chopper. The time-resolved measurements are achieved by modifying the time delay $\tau_\text{delay}$ between the pump and probe beams. This enables the detection of optical changes versus time with sub-picosecond resolution. A cryostat with a superconducting magnet is utilized to control the applied magnetic field and sample temperature. We apply magnetic fields which point in and out of the sample plane in the $<110>$ crystal direction and refer to them as in-plane and out-of-plane measurements. The in-plane and out-of-plane TR-MOKE signals reflect thermally-induced changes in magnetization perpendicular (dm/dt $\perp H$) and parallel (dm/dt $\parallel H$) to the applied magnetic field, respectively.
%% experimental results

%Figure 1
In Figs. \ref{fig:Figure1}(b) and (c), in- and out-of-plane TR-MOKE signals are presented for the field-aligned (FA), conical, helical, and paramagnetic phase. Following optical pump beam excitation at zero time, we observe a rapid rise in the TR-MOKE signal indicating thermal alteration of magnetic order by laser heating \cite{beaurepaire1996ultrafast}. Subsequently, the signal decreases as the magnetic state recovers. In the FA and conical phase, recovery manifests as collective precessional dynamics observed in the in-plane TR-MOKE measurements in Fig. \ref{fig:Figure1}(c).
As collective dynamics predominantly occur in the sample plane perpendicular to the applied magnetic field for the FA and conical phase, the out-of-plane TR-MOKE signals exhibit a monotonous decay, consistent with prior studies on chiral magnets \cite{koralek2012observation, kalin2022optically}. In the conical phase, the out-of-plane TR-MOKE signal decays on a much faster timescale compared to the FA phase, with a small overshoot to negative values. Precessional remagnetization behavior is observed for the helical phase in both the in- and out-of-plane TR-MOKE signals, attributed to the alignment of the helical phase with the direction of magnetically easy axes. 
An analysis of the oscillation frequency as a function of the applied field shows that the observed precessional dynamics exhibit the characteristic dispersion of the spin eigenmodes of the different magnetic phases of MnSi \cite{schwarze2015universal}, as illustrated in Fig. \ref{fig:Figure1}(d). 
For the paramagnetic phase, only the out-of-plane TR-MOKE signal for $B>0\,\text{T}$ is non-zero. Here, we observe an increase of $\Delta \phi_k$ with the applied magnetic field, see Fig. \ref{fig:Figure1}(e), indicating field-induced spin alignment. Overall, the TR-MOKE measurements indicate a thermally induced alteration of the magnetic order by laser heating. Thereby, the different magnetic phases of MnSi can be unambiguously identified using characteristic signatures within the TR-MOKE signal.

\begin{figure}[!]
\includegraphics{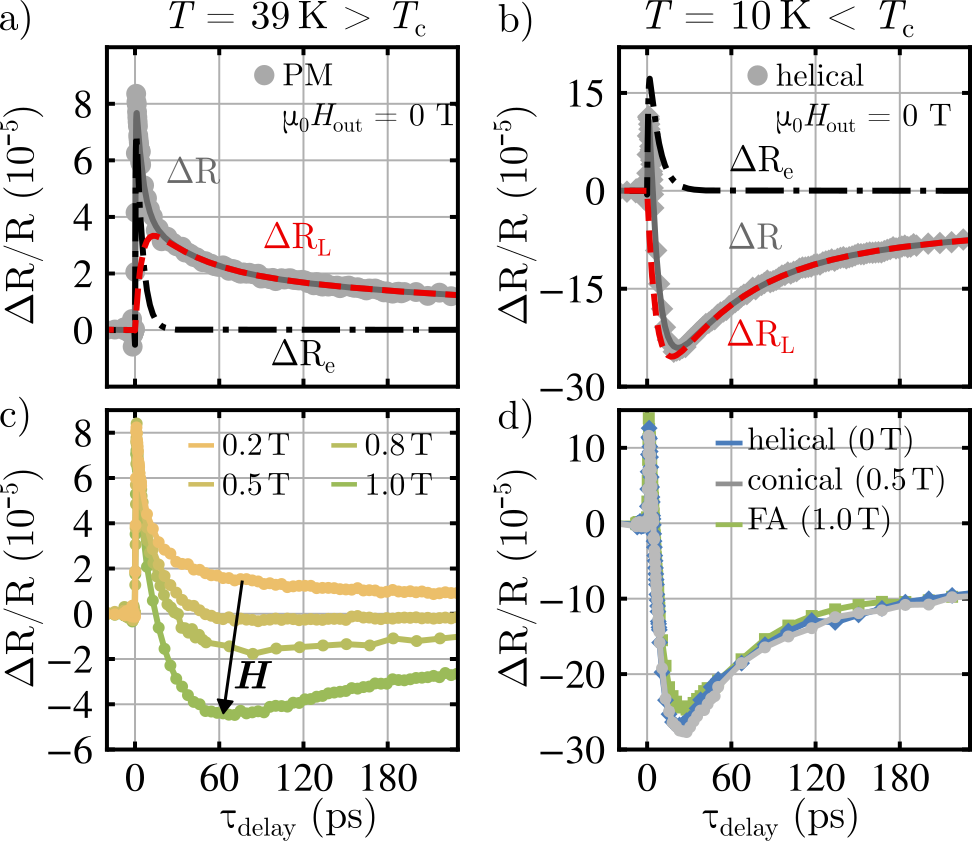}
\caption{\label{fig:Figure2} Transient reflectivity of the different magnetic states of MnSi. (a) Paramagnetic phase at ${T=39\,\text{K}}$ and ${\mu_0 H_\text{out}= 0\,\text{T}}$ shown with fit to the experimental data using the model of Eq. \ref{Eq:bipolar_transient}. b) Helical phase at ${T=10\,\text{K}}$ and ${\mu_0 H_\text{out}= 0\,\text{T}}$ shown with fit to the experimental data using the model of Eq. \ref{Eq:bipolar_transient}. (c) Paramagnetic phase at ${T=39\,\text{K}>T_\text{c}}$ and various magnetic fields, which point out of the sample plane along the $<110>$ direction. (d) States with long-range magnetic order at low temperatures ${T=10\,\text{K}}$ namely helical phase at $0\,\text{T}$, conical phase at $0.5\,\text{T}$ and field-aligned phase at $1.0\,\text{T}$. The magnetic field is pointing out of the sample plane along the $<110>$ direction.
}
\end{figure}

%Figure 2
%temperature dependence
Next, we study the transient reflectivity of the different magnetic phases of MnSi. In Figs. \ref{fig:Figure2}(a) and (b) the transient reflectivity at zero magnetic field well above and below $T_\text{c}$, i.e., in the paramagnetic and helical phase, is shown. For the paramagnetic phase, we observe a rapid increase of $\Delta R/R$ upon application of the pump pulse at ${t=0\,\text{ps}}$. After this the transient reflectivity shows a exponential decay as the signal gradually returns to its initial state, resulting in an unipolar transient reflectivity signal. In contrast, in the helical phase the rapid increase of the reflectivity signal is followed by a zero-crossing and a large negative transient, leading to a bipolar temporal evolution of the thermally induced reflectivity change \cite{comment1}. Thus, the transient reflectivity at zero magnetic field shows a different behavior well above and below the critical temperature.

%% magnetic field dependence: paramagnetic state
In Figs. \ref{fig:Figure2}(c) and (d), we show transient reflectivity measurements for different magnetic fields for $T>>T_c$ and $T<T_c$.
Well above $T_\text{c}$, we observe a transition from the unipolar transient to the bipolar signal with increasing magnetic field. Here, the applied magnetic field forces an alignment of magnetic moments in the paramagnetic phase. In the magnetically-ordered states below $T_\text{c}$, however, we observe independent of the applied magnetic field and magnetic configuration a bipolar signal, as shown by the comparison of the transient reflectivity of the helical, conical and field-aligned phase in Fig. \ref{fig:Figure2}(d). Accordingly, the bipolar reflectivity signal occurs independently of the exact magnetic state. Thereby, the bipolar reflectivity transient is observed independent of the magnetic field polarity and direction, i.e., out-of- and in-plane magnetic fields, as well as the polarization of the probe beam (not shown). Together with the absence of contributions attributed to precessional dynamics in the reflectivity transient, especially at $0\,\text{T}$ in the helical phase, this rules out magneto-optical effect as cause of the bipolar reflectivity. At the same time it confirms that the TR-MOKE measurements in Fig. \ref{fig:Figure1} represent the magnetization dynamics and not the change of reflectivity \cite{kampfrath2002ultrafast}.

%-------------

% model
To analyze the transient reflectivity of MnSi further, we apply the conventional phenomenological model for metals and assume that the reflectivity is proportional to the electron and lattice temperature \cite{djordjevic2006comprehensive, liu2021differentiating}. Thus $\Delta R/R$ is a superposition of an electronic $\Delta R_\text{e}/R$ and a lattice reflectivity transient $\Delta R_\text{L}/R$ 
\begin{equation}
\begin{split}
\frac{\Delta R(\tau_{\text{delay}})}{R}= \frac{\Delta R_\text{e}(\tau_{\text{delay}})}{R} + \frac{\Delta R_\text{L}(\tau_{\text{delay}})}{R}= A\cdot \Delta T_\text{e}(\tau_{\text{delay}}) + B\cdot \Delta T_\text{L}(\tau_{\text{delay}}).\label{Eq:bipolar_transient}
\end{split}
\end{equation}
The term $\Delta R_\text{e}/R$ describes the reflectivity change due to the increase of the electron temperature $T_\text{e}$ by laser heating and accounts for the laser-excited electrons and their thermalization. The second term  $\Delta R_\text{L}/R$ models the transient reflectivity contribution stemming from an increase in lattice temperature $T_\text{L}$ and covers the change of the optical reflectivity due to thermal expansion of the lattice. Thereby, $A$ and $B$ describe the proportionality of $\Delta R_\text{e}/R$ and $\Delta R_\text{L}/R$ to the change in electron and lattice temperature, respectively. Both $A$ and $B$ can be positive or negative depending on how the optical transition rates and thus the optical reflectivity are modulated by the change in the electron and lattice distribution \cite{djordjevic2006comprehensive}.
The key components of the transient electron $\Delta T_\text{e}$ and lattice $\Delta T_{\text{L}}$ temperature are exponential rise and decay functions. They are given by
\begin{gather}
\Delta T_\text{e}(\tau_{\text{delay}}) =\Delta T_\text{e,max}\cdot [1-\exp(-\tau_{\text{delay}}/\tau_\text{ee})]\cdot\exp(-\tau_{\text{delay}}/\tau_\text{eL})\cdot\Theta(\tau)\\
\Delta T_\text{L}(\tau_{\text{delay}}) = \Delta T_\text{L,max}\cdot [1-\exp(-\tau_{\text{delay}}/\tau_\text{eL})]\cdot[\beta_1 \cdot\exp(-\tau_{\text{delay}}/\tau_\text{th})+
\\ \beta_2\cdot \exp(-\tau_{\text{delay}}/\tau_\text{th,2})]\cdot\Theta(\tau). \notag
\end{gather}
Here, $\Delta T_\text{e,max}$ and $\Delta T_\text{L,max}$ are the maximum change of the electron and lattice temperature due to laser heating, which we estimate in a framework based on the 3T model \cite{supplement}. The time constant $\tau_{\text{ee}}$ describes the rise time of the electron reflectivity transient, $\tau_{\text{eL}}$ is the electron-lattice thermalization, $\tau_{\text{th}}$ is the diffusion time constant and $\tau_\text{th,2}$ is a second decay time of $\Delta R_\text{L}$. The $\tau_{\text{ee}}$ is determined by the time the electron system needs to equilibrate in energy and form a Fermi-Dirac distribution \cite{djordjevic2006comprehensive}. Further, $\tau_{\text{ee}}$ is influenced by the laser pulse width, because it determines the time to deposit the laser energy into the electron system \cite{djordjevic2006comprehensive}. The second decay time $\tau_\text{th,2}$ accounts for the double exponential decay of the negative transient observed in Figs. \ref{fig:Figure2}(b)-(d). 
The parameters $\beta_{i}$ represent the amplitudes of the exponential decay terms of $\Delta T_\text{L}$, ranging between 0 and 1. The term $\Theta(\tau)$ is an error function and describes the finite width of the probe pulse and the start of excitation at $t = 0$. %which we determine in using the 2T-model and specific heat data \cite{supplement}
In the analysis, we assume that the thermalization of non-thermal electrons and lattice is comparable or smaller than the laser pulse width. Therefore, we do not consider non-thermal dynamics of electrons and the lattice \cite{PhysRevX.6.021003}.

%The first term proportional to α represents the electronic response, initially determined by the electron- electron thermalization (with time constant τee) and then decaying by energy transfer to the lattice with the characteristic electron-phonon relaxation time τep. The second term, proportional to β, accounts for the lattice heating and thus rises with the same time constant τep; the additional relaxation parameter is due to heat diffusion outside the irradiated area τth

In Figs. \ref{fig:Figure2}(a) and (b) fits of the phenomenological model to the unipolar and bipolar reflectivity transient are shown. In both cases, we obtain excellent agreement of model and the data. Remarkably, the electronic $\Delta R_\text{e}/R$ and lattice reflectivity transient $\Delta R_\text{L}/R$ are both positive for the unipolar signal, while $\Delta R_\text{L}/R$ is negative in the case of the bipolar transient.
The transition from a unipolar to a bipolar reflectivity transient can therefore be related to a change in polarity of $\Delta R_\text{L}/R$.

\begin{figure}
\includegraphics{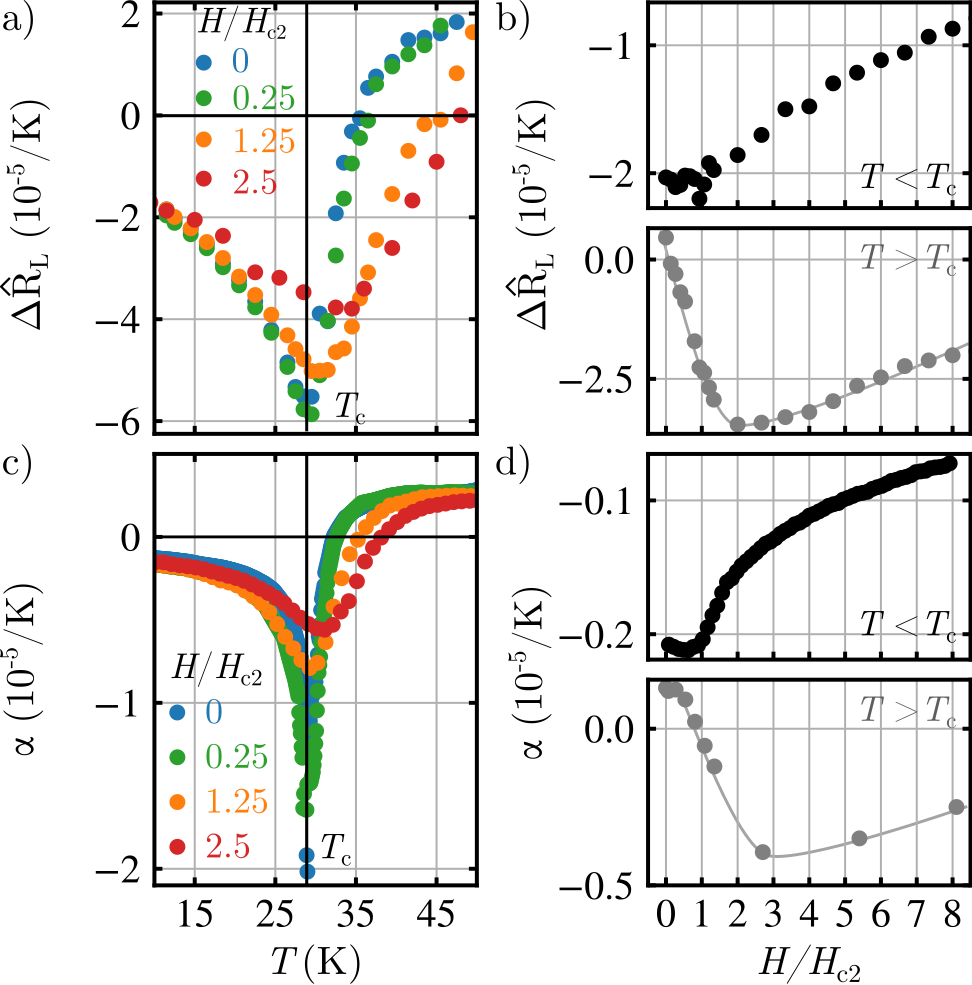}
\caption{\label{fig:Figure3}(a) Temperature dependence of the amplitude of the lattice reflectivity transient $\Delta \hat{R}_\text{L}$ at four different magnetic field values. (b) Magnetic field dependence of $\Delta \hat{R}_\text{L}$ below ($10\,\text{K}$, top) and above ($35\,\text{K}$, bottom) the critical temperature. 
(c) Coefficient of linear thermal expansion $\alpha$ of MnSi from Ref. \cite{stishov2007magnetic} as a function of temperature for four different magnetic field values. (d) Magnetic field dependence of $\alpha$ at $10\,\text{K}<T_c$ from Ref. \cite{petrova2016thermal} (top) and at $35\,\text{K}>T_c$ from Ref. \cite{stishov2007magnetic}.}
\end{figure}

% Figure s3(a) and (b): Estimation of thermal expansion 
We further investigate the polarity of $\Delta R_\text{L}/R$ by analyzing its amplitude as a function of temperature and applied magnetic field. According to Eq. 1-3 the amplitude of $\Delta R_\text{L}/R$ is equal to $B\cdot\Delta T_\text{L,max}$. We exclude the temperature dependence of $\Delta T_\text{L,max}$ \cite{supplement} in our analysis by considering $\Delta \hat{R}_\text{L}=B$ in the following.
As illustrated in Fig. \ref{fig:Figure3}(a) for various magnetic fields, the amplitude of the transient reflectivity $\Delta \hat{R}_\text{L}$ changes sign with temperature.
For $T<T_c$, $\Delta \hat{R}_\text{L}$ is negative and decreases at higher temperatures. At $T_\text{c}$, $\Delta \hat{R}_\text{L}$ shows its minimum and increases versus temperature until it turns positive. Subsequently, the positive $\Delta \hat{R}_\text{L}$ rises with temperature. Further, we observe a shift of the minimum $\Delta \hat{R}_\text{L}$ to higher temperatures with increasing magnetic field. At the same time, zero crossing of $\Delta \hat{R}_\text{L}$ occurs for larger temperatures upon increase of the magnetic field.

The magnetic field dependence of $\Delta \hat{R}_\text{L}$ is presented in Fig. \ref{fig:Figure3}(b). For $T<T_c$, $\Delta \hat{R}_\text{L}$ varies only slightly with magnetic field up to $H_\text{c2}$ and increases for higher values in the field-aligned phase (see upper panel in Fig. \ref{fig:Figure3}(b)). Thereby, it remains negative in the entire magnetic field range. At $T>T_c$, we observe a different behavior of the lattice transient reflectivity, as shown in the lower panel in Fig. \ref{fig:Figure3}(b) for $35\,\text{K}$. Here, $\Delta \hat{R}_\text{L}$ first decreases for increasing magnetic field and then increases again, possibly due to a transition from the paramagnetic to the field-aligned phase. Thereby, the sign of $\Delta \hat{R}_\text{L}$ changes from positive to negative for small magnetic fields at $T>35\,\text{K}$, as shown as well in Fig. \ref{fig:Figure3}(a). Overall, we observe a different magnetic field dependence of $\Delta \hat{R}_\text{L}$ for $T<T_c$ and $T>T_c$.

% Figure 3(c) and (d)
According to the conventional model applied above, the lattice contribution to transient reflectivity is due to the changes in the optical properties caused by thermal expansion of the lattice \cite{eesley1983observation, liu2021differentiating, djordjevic2006comprehensive}.
With the help of the amplitude analysis of $\Delta R_\text{L}/R$, we can compare the transient reflectivity measurements with thermal expansion data.
%The comparison of a transient signal and a static measurement is applicable here, as the maximum signal of $\Delta R_\text{L}$ appears after thermal equilibration of electrons, lattice, and spin.
For the comparison, we take the coefficient of linear thermal expansion $\alpha = (\text{1/L}_0)\text{dL/dT}$ measured with a capacity dilatometer on a comparable MnSi sample from Ref. \cite{stishov2007magnetic} and \cite{petrova2016thermal} as a measure of thermal expansion. Here, $L_0$ is the sample length at $300\,\text{K}$ and ${\text{dL/dT}}$ is the change of sample length with temperature. For the data from Ref. \cite{petrova2016thermal} we approximate $\alpha$ by $(\text{1/L}_0) (\Delta L(T_2)-\Delta L(T_1))/(T_2-T_1)$ using measurements of the linear thermal expansion $\Delta L/L_0$ at different temperatures. %Thereby, $T_2-T_1 \approx \Delta T_\text{s,max}$.}

In Figs. \ref{fig:Figure3}(c) and (d) the coefficient of linear thermal expansion $\alpha$ as a function of temperature and applied magnetic field is shown.
Like the amplitude of the lattice reflectivity transient, $\alpha$ peaks at $T_\text{c}$ and changes sign with increasing temperature. Note, that the sharp peak at $T_\text{c}$ seen in $\alpha$ for ${H<H_\text{c2}}$ is not observed for $\Delta \hat{R}_\text{L}$, most likely due to larger temperature increases from laser heating and the averaging of the temperature-dependent signal in the transient reflectivity data. Further, $\alpha$ has a very similar dependence on the magnetic field as $\Delta \hat{R}_\text{L}$, since it shows the distinct variations with field above and below $T_\text{c}$. %The almost constant value of $\alpha$ in the helical and conical phase is due to the fact that the magnetic order parameter is not directly coupled to the magnetic field \cite{stishov2007magnetic}. In the field-aligned state ($H/H_\text{c2}>1$), on the other hand, $\alpha$ scales with the magnetic field, as the magnetic field increases the magnetic order in this state.
The similarity of $\Delta \hat{R}_\text{L}$ and $\alpha$ indicates that the shift in polarity of $\Delta R_\text{L}/R$ can be explained by the temperature and magnetic field dependence of thermal expansion. It's noteworthy that specific heat, thermal expansion, and the temperature derivative of resistivity exhibit a unified temperature dependence \cite{stishov2008heat}. However, the sign change only observed for the thermal expansion is the driving force for the polarity reversal of $\Delta R_\text{L}/R$ and the transition from unipolar to bipolar reflectivity transient.

\begin{figure}
\includegraphics{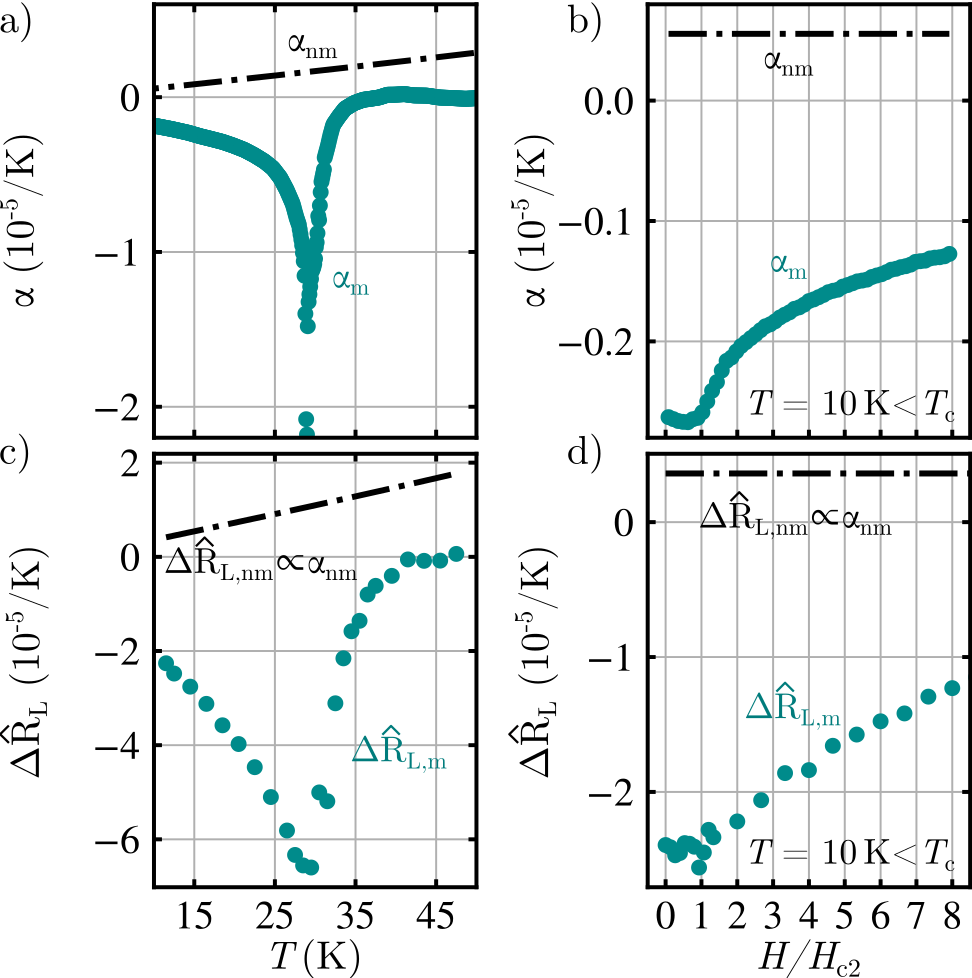}
\caption{\label{fig:Figure4} (a) and (b) Magnetic (solid turquoise line) and non-magnetic (dashed black line) coefficient of linear thermal expansion of MnSi at $0\,\text{T}$ as a function of temperature (a) and at $10\,\text{K}$ as a function of magnetic field (b). (c) and (d) Temperature and magnetic field dependence of the magnetic (solid turquoise line) and non-magnetic (dashed black line) contributions to the lattice transient reflectivity at $0\,\text{T}$ (c) and $10\,\text{K}$ (d), respectively.
}
\end{figure}

%figure 4
To analyze the polarity change of $\Delta R_\text{L}/R$ with temperature in more detail, we divide the thermal expansion of MnSi in its non-magnetic and magnetic contributions. The first contribution is due to thermal expansion caused by the excitation of conduction electrons as well as phonons and the latter contribution is due to the magneto-volume effect. To that end, we use the method proposed in Ref. \cite{matsunaga1982magneto}. In this approach, it is assumed that the non-magnetic coefficient of linear thermal expansion is $\alpha_\text{nm}=k_e T+k_pT^3$ and not dependent on the magnetic field. This is because thermal expansion due to the excitation of conduction electrons and phonons are proportional to $T$ and $T^3$, respectively, and are not affected by magnetic field changes. We extract $k_e$ and $k_p$ by a least square fit to $\alpha$. Further, we subtract $k_e T+k_pT^3$ from $\alpha$ to obtain the magnetic coefficient of linear thermal expansion $\alpha_\text{m}$. In  Figs. \ref{fig:Figure4}(a) and (b), $\alpha_\text{m}$ and $\alpha_\text{nm}$ are shown exemplarily for $0\,\text{T}$ as a function of temperature and at $10\,\text{K}$ as a function of the magnetic field, respectively. As known from literature \cite{matsunaga1982magneto}, the negative $\alpha_\text{m}$ dominates $\alpha$ for $T<T_c$, peaks at $T_\text{c}$ and is zero for $T>>T_c$. Further, $\alpha_\text{m}$ and thus magneto-volume effect are the reason for the pronounced magnetic field dependence of $\alpha$.
%proximity to phase transitions

%Note here, that the lattice and phonon contributions could not be extracted in the case of magnetic field dependent $\alpha$ data from Ref. \cite{petrova2016thermal}, as the required temperature dependent data is missing. In this case, we use the corrections determined from the temperature dependent measurements from Ref. \cite{stishov2007magnetic} at $0\,\text{T}$.
Using the analysis of $\alpha$, we separate the magnetic $\Delta \hat{R}_\text{L,m}$ and non-magnetic $\Delta \hat{R}_\text{L,nm}$ contribution to the lattice transient reflectivity. Thereby, we assume that $\Delta \hat{R}_\text{L,nm}$ is proportional to $\alpha_\text{nm}$. Further, by subtracting $\Delta \hat{R}_\text{L,nm}$ from $\Delta \hat{R}_\text{L}$ we extract the magnetic contribution to the lattice transient reflectivity $\Delta \hat{R}_\text{L,m}$, which is caused by thermal expansion due to the magneto-volume effect. $\Delta \hat{R}_\text{L,m}$ and $\Delta \hat{R}_\text{L,nm}$ are shown in Figs. \ref{fig:Figure4}(c) and (d) as a function of temperature and magnetic field, respectively. The lattice transient reflectivity is dominated by the negative $\Delta \hat{R}_\text{L,m}$ at low temperatures, while at $T>>T_c$ the positive $\Delta \hat{R}_\text{L,nm}$ determines the lattice transient reflectivity. 
Thus, the bipolar reflectivity transient of Fig. \ref{fig:Figure2}, which we associated with a negative $\Delta R_\text{L}/R$ in Fig. \ref{fig:Figure3}, is due to magneto-volume effect dominating the thermal expansion below and in proximity to $T_\text{c}$. Thereby, the magneto-volume effect changes the optical reflectivity in an opposite manner than the non-magnetic thermal expansion, due to its inverse effect on the lattice. With that, this study shows that $\alpha_\text{m}$ and thus the magneto-volume effect must be carefully considered when interpreting the transient reflectivity of magnetic materials.
%Thus, the temperature increase and by that the alteration of magnetic order induced by thermal laser excitation, gives rise to a volume change due to the magneto-volume effect. This, in turn, affects the optical properties, such as interband or intraband scattering rates, of the MnSi sample, which we probe by the transient reflectivity.}

%Energy shifts are linear for small changes in the lattice spacing \cite{winsemius1976temperature}, % dielelectric function influence by electron-phonon coupling \cite{xu2017role}

% Summary
In conclusion, this study explores the transient reflectivity of MnSi under femtosecond alteration of long-range magnetic order. 
The transient reflectivity in MnSi transforms from a unipolar behavior in the paramagnetic phase for $T>>T_c$ to a bipolar signal in the states exhibiting magnetic long-range order for $T<T_c$. By dividing the transient reflectivity in electron and lattice contribution, our findings explain the bipolar response by the magneto-volume effect dominating the thermal expansion of MnSi. Overall this study shows that magneto-volume effect can influence the transient reflectivity and needs to be considered for a correct estimation of the transient temperature response after femtosecond laser excitation. Our work shows that the typical framework of  electron and lattice transient reflectivity is valid even when the magneto-volume effect dominates the thermal expansion. This paves the way to study the time constants of the magneto-volume effect by transient reflectivity measurements, which might give further insights into the dynamics of magneto-elastic coupling.
% This is important for ..

%We evaluate if the standard model for transient reflectivity is sufficient to describe the transient reflectivity of a material with strong magneto-volume effect. In this framework, we observe a polarity change of the lattice transient reflectivity.

%- additional contribution to transient reflectivity
%- when accounting for this additional contribution the timescales of thermal relaxation can be determined more precisely, but might also give insights in further timescales such as the magnetoelastic coupling

%By that, we introduce a tool to analyze the characteristic timescales of ultrafast magneto-volume effect and magnetoelastic coupling.

%Our finding are important to interpret . The change of reflectivity due to the magneto-volume effect might affect time-resolved magneto-optical effects, a magnetization sensitive measurement, as well.

%Therefore, it becomes essential to disentangle the dynamics of spin, electrons, and lattice in order to interpret magnetization dynamics adequately, particularly in the context of transient magneto-optical Kerr effect (TR-MOKE) measurements.

\begin{acknowledgments}
We wish to thank Marcelo Jaime for support and stimulating discussions.\\
\textbf{Funding:} This work was supported by the European Metrology Research Programme (EMRP) and EMRP participating countries under the European Metrology Programme for Innovation and Research (EMPIR) Project No. 17FUN08-TOPS Metrology for topological spin structures. In part, this study has been funded by the Deutsche Forschungsgemeinschaft (DFG, German ResearchFoundation) under TRR80 (From Electronic Correlations to Functionality, Project No.\ 107745057, Project E1), TRR360 (Constrained Quantum Matter, Project No.\ 492547816, Projects C1 \& C3), SPP2137 (Skyrmionics, Project No.\ 403191981, Grant PF393/19 and Grant SCHU 2250/8-1), the excellence cluster MCQST under Germany's Excellence Strategy EXC-2111 (ProjectNo.\ 390814868) and EXC-2123 QuantumFrontiers (ProjectNo. 390837967). Financial support by the European Research Council (ERC) through Advanced Grants No.\ 291079 (TOPFIT) and No.\ 788031 (ExQuiSid) is gratefully acknowledged.
\textbf{Author contributions:} MB and HWS initiated the study. AB and CP grew the crystal. JK performed the experimental study under the supervision of MB and HF. The data analysis was performed by JK. All authors discussed the results. JK wrote the paper with comments of all authors. 
\textbf{Competing interests:} The authors declare that they have no competing interests. 
\textbf{Data availability}: All data needed to evaluate the conclusions of the paper are present in the paper and the supplemental material. 
\end{acknowledgments}

%\section*{Data Availability Statement}

\nocite{*}
\bibliography{References.bib}% Produces the bibliography via BibTeX.

\end{document}